\documentclass[12pt,nohyper,letterpaper]{JHEP3}         
\usepackage{amssymb,amsfonts,amstext}

\usepackage{epsfig}

\def\be{\begin{eqnarray}}
\def\ee{\end{eqnarray}}

\newcommand\para{\paragraph{}}

\def\Dslash{\,\,{\raise.15ex\hbox{/}\mkern-12mu D}}
\def\Dbarslash{\,\,{\raise.15ex\hbox{/}\mkern-12mu {\bar D}}}
\def\delslash{\,\,{\raise.15ex\hbox{/}\mkern-9mu \partial}}
\def\delbarslash{\,\,{\raise.15ex\hbox{/}\mkern-9mu {\bar\partial}}}
\def\pslash{\,\,{\raise.15ex\hbox{/}\mkern-9mu p}}
\def\calDslash{\,\,{\raise.15ex\hbox{/}\mkern-12mu {\cal D}}}

\def\lae{\mathrel{\mathop{\smash{\lower .5 ex \hbox{$\stackrel<\sim$}}}}}
\def\lae{\mathrel{\mathop{\smash{\lower .5 ex \hbox{$\stackrel>\sim$}}}}}

\def\theequation{\thesection.\arabic{equation}}

\def\Dslash{\,\,{\raise.15ex\hbox{/}\mkern-13mu D}}
\def\Dbarslash{\,\,{\raise.15ex\hbox{/}\mkern-12mu {\bar D}}}
\def\delslash{\,\,{\raise.15ex\hbox{/}\mkern-10mu \partial}}
\def\delbarslash{\,\,{\raise.15ex\hbox{/}\mkern-9mu {\bar\partial}}}
\def\pslash{\,\,{\raise.15ex\hbox{/}\mkern-11mu p}}
\def\qslash{\,\,{\raise.15ex\hbox{/}\mkern-9mu q}}
     \def\kslash{\,\,{\raise.15ex\hbox{/}\mkern-11mu k}}
\def\eslash{\,\,{\raise.15ex\hbox{/}\mkern-9mu \epsilon}}
\def\calDslash{\,\,{\rais.15ex\hbox{/}\mkern-12mu {\cal D}}}

\title{Tunneling $D0$-branes}
\author{Chris Pedder\\
Department of Applied Mathematics and Theoretical Physics, \\
University of Cambridge, UK\\{\tt cjp46@cam.ac.uk}}

\abstract{In the $D0$-$D4$-brane system, $D0$-branes do not tunnel. Instead, they form bound states with the $D4$-brane, whose ground states are exact. However, the $D0$-brane quantum mechanics contains a BPS instanton. To what does this solution correspond?  We find that such a tunneling solution provides non-perturbative corrections to the Berry phase connection for the first-excited states as the $D4$-branes are moved adiabatically. We compute this connection for the first four excited states, and show that it gives an emergent $SO(5)$ connection described previously by Tchrakian.
}

\begin{document}
\pagestyle{plain} \setcounter{page}{1}
\newcounter{bean}
\baselineskip16pt

\section{Introduction}

The purpose of this paper is to study the tunneling of $D0$-branes in the $D0$-$D4$-brane system. We study the particular case of a $D0$-brane living between two, separated $D4$-branes. Ordinarily, we would expect there to be no force between the two different types of brane, however we introduce a $NS$-$NS$ $B$-field which causes the $D0$-brane to be attracted to the $D4$-branes. There then exist two possible vacua of the system, with the $D0$-brane nestled inside one or other of the $D4$s. In this static arrangement, the ground states are exact, so there is no tunneling solution between them.
\para
We describe the system by means of the quantum mechanics of the $D0$-brane, which contains BPS kink solutions that tunnel from one $D4$-brane vacuum to the other. Such quantum-mechanical instantons have provided interesting contributions to physical quantities in the past. They are crucial to the derivation of the strong morse inequalities in $(1,1)$-supersymmetric theories ~\cite{Witten:1982im}, and give off-diagonal contributions to the ground-state Berry connection in $(2,2)$- supersymmetric theories ~\cite{geophase}. We wish to answer the question; to what do the quantum mechanical kinks in our $(4,4)$-supersymmetric system contribute?
\para
We will show that the answer to this question is that they contribute to the Berry phase connection for the system. This arises in the following way. We take our pair of $D4$-branes and absorb the $D0$-brane in one of them. We then move the $D4$-branes slowly around a closed path in the space transverse to them (this is ${\mathbb{R}}^5$ since $D4$-branes have a five-dimensional worldvolume in ten-dimensions), so that they come back to where they started. We then look to see in which state the $D0$-brane now lives.
\para
We find that the $D0$-brane ground states do not change under this evolution. However, as we shall see, there are
two particular first-excited states above each of the vacua that do tunnel as we move the $D4$ branes. Since they
are degenerate in energy, we expect these $2\times 2=4$ states $|a_i \rangle \; , \; i=1,\dots,4$ to undergo a
holonomy which depends on the path taken by the $D4$-brane. We let the $D4$-branes be separated by a distance
$2 {\vec{M}}$, so that the states change according to
\be
| a_i \rangle \rightarrow P \exp \left( -i \oint ({\omega}_{\mu})_{ij}({\vec{M}}) \, dM^{\mu} \right) |a_j \rangle.
\ee
under adiabatic motion of the $D4$-brane. This holonomy is given in terms of a non-Abelian $u(4)$ \emph{Berry
phase connection}, ${\omega}_{\mu}$, which we aim to compute. The Berry phase connection may be computed from the
states $|a_i \rangle$ as
\be
({\omega}_{\mu})_{ij} = i \langle a_i| \frac{\partial}{\partial M^{\mu}}| a_j \rangle .
\ee
It will transpire that this connection has interesting geometrical properties. We will show that before including
tunneling effects, the connection ${\omega}_{\mu}$ takes the form
\be
{\omega}_{\mu} = \left( \begin{array} {cc} A^{\star}_{\mu} & 0 \\ 0 & A_{\mu} \end{array} \right).
\ee
$A_{\mu}$ is an $su(2)$ connection over ${\mathbb{R}}^5$ that is often referred to as the \emph{Yang monopole} ~\cite{yang}, and $A^{\star}_{\mu}$ is its complex conjugate, the \emph{Yin monopole}. The Yang monopole connection $A_{\mu}$ obeys the topological condition that
\be
c_2(A) = \int F \wedge F = -1,
\ee
where $F=dA$, the field-strength, whereas the Yin monopole has $c_2=+1$. We will describe the total connection ${\omega}_{\mu}$ in more detail in section 3.
\para
The main result of this paper concerns BPS instantons, which give off-block-diagonal contributions to the original connection ${\omega}_{\mu}$. Remarkably, when we include these corrections our Berry holonomy is found to live in $SO(5)$, and not $U(4)$ as we would have expected. It is not understood how this change in structure occurs. The $SO(5)$-monopole has been studied already, notably by Tchrakian ~\cite{Tchrakian:2006nn}. We will refer to it as the ``Tchrakian monopole''. It can be understood as the $SO(5)$ generalization of the Yang monopole in much the same way as the 't Hooft-Polyakov monopole is a generalisation of the Dirac monopole. We will later compare our $SO(5)$-monopole with those appearing in the literature.
\para
The structure of the paper is as follows. In section 2, we introduce the $(4,4)$ quantum mechanics of interest, paying particular attention to the parameters we may introduce. We describe the two vacua of the system, and the spectrum of excited states living above them. We show that there can be no BPS kink correction to either the ground states, or the first excited states of the static system.
\para
Section 3 covers the possibility of adiabatic motion of the $D4$-branes. We discover that under such motion, it is the first excited states of the system that acquire a non-Abelian Berry phase connection at the perturbative level. We then investigate the possibility of non-perturtbative corrections to the Berry phase. We find the Bogomolnyi equations that describe the BPS kinks in the $(4,4)$ quantum mechanics, and look for zero-mode solutions to the Dirac equations governing the fermions. We then compute the Berry phase overlaps explicitly, and use our knowledge of the fermi zero-modes to discover what form these corrections may take. We explicitly rotate this non-perturbative connection to a spherically symmetric gauge, where the total connection takes the form of the Tchrakian $SO(5)$-monopole. We then make some comments on the full connection.
\para
Various computational complexities are relegated to a pair of appendices.
\para
\section{$D0$-$D4$ Quantum Mechanics}

In this section we will describe the quantum mechanics that characterises the behaviour of a single $D0$-brane in the background of two $D4$-branes in type IIA string theory. The model we study exhibits ${\mathcal{N}} = (4,4)$ supersymmetry, and descends from the $\mathcal{N} = 1$ theory in $d=5+1$ dimensions. The massless representations include the familiar hypermultiplet and vector multiplet, understood to be excitations of the $D0$-$D4$ strings, and the $D0$-brane respectively. We will neglect the motion of the $D0$-brane parallel to the $D4$-branes, which would give a $16$-fold degeneracy to the states that we find, but is otherwise trivial. The automorphism group of the superalgebra is
\be
R=Spin(5)\times SU(2)_R.
\ee
$Spin(5)$ is the unbroken part of the $SO(9,1)$ Lorentz symmetry that describes the geometric rotations of the space transverse to the branes, and $SU(2)_R$ is part of the $SO(4) = SU(2)_L \times SU(2)_R$ symmetry of the worldvolume of the $D4$-branes.
\para
We will work with a Lagrangian which takes the form
\be
L = L_{vector} + L_{hyper} + L_{Yuk}.
\label{lagrangian}
\ee
\para
The vector multiplet consists of the vector ${\vec{X}}$ which parameterises the ${\mathbb{R}}^5$ transverse to the $D4$-branes, and transforms in the ({\bf 5}, 1) of $R$. It also contains ${\Lambda}_{\alpha} \; \alpha = 1,\dots,4$, a four-component complex spinor that transforms in the $({\bf 4}, {\bf 2})$ of $R$. The first term in (\ref{lagrangian}) describes the motion of the $D0$-brane transverse to the $D4$s.
\be
L_{vector} = \frac{1}{2g^2} ({\dot{\vec{X}}}^2 + 2i {\bar{\Lambda}} {\dot{\Lambda}}).
\ee
In string theoretic language, the distance between the $D0$-brane and the $D4$-branes is given by $2 \pi {\alpha}^{\prime} {\vec{X}}$ and the coupling $g^2 = g_s /(2\pi)^2 {{\alpha}^{\prime}}^{3/2}$. The decoupling limit of the quantum mechanics is $g_s \rightarrow 0$ and ${\alpha}^{\prime} \rightarrow 0$ with ${\vec{X}}$ and $g^2$ fixed. The mass of the $D0$-brane is then $M_{D0} = 1/g_s \sqrt{{\alpha}^{\prime}}$.
\para
The hypermultiplets describing the $D0$-$D4$ strings consist of a pair of complex scalars $({\phi}_i, {\tilde{\phi}}_i )$ which can be paired up into the $SU(2)_R$ doublet $\omega = ({\phi}_i, {\tilde{\phi}}^{\dagger}_i )^T$. They also contain a complex spinor ${\Psi}_{i \alpha}$ which transforms in the $({\bf 4}, 1)$ of $R$. The hypermultiplets are governed by the Lagrangian
\be
L_{hyper} &= \sum_{i=1}^2 |{\mathcal{D}}_t {\phi}_i|^2 + |{\mathcal{D}}_t {\tilde{\phi}}_i|^2
+i {\bar{\Psi}}_i {\mathcal{D}}_t {\Psi}_i - {\vec{X}}^2 (|{\phi}_i|^2 + |{\tilde{\phi}}_i|^2) \nonumber \\
&\;\;\; - \frac{g^2}{2} (\sum_{i=1}^2 |{\phi}_i|^2 - |{\tilde{\phi}}_i|^2)^2 - 2g^2|\sum_{i=1}^2 {\tilde{\phi}}_i {\phi}_i|^2 .
\label{hyper}
\ee
The ${\vec{X}}^2 |{\phi}_i|^2$ terms in (\ref{hyper}) describe the mass of a string stretched a distance ${\vec{X}}$ between the $D0$-brane and a $D4$-brane.
\para
The Yukawa terms are given by
\be
L_{Yuk} = \sum_{i=1}^2 -{\bar{\Psi}}_i ({\vec{X}} \cdot {\vec{\Gamma}}) {\Psi}_i + (\sqrt{2} {\bar{\Psi}}_{i \alpha} ({\phi}_i {\Lambda}_{\alpha} + {\tilde{\phi}}^{\dagger}_i J_{\alpha}^{\; \beta} {\Lambda}_{\beta}^{\star}) +\text{h.c.}),
\ee
where the matrices ${\vec{\Gamma}}$ obey the $SO(5)$ Clifford algebra $\{ {\Gamma}_a , {\Gamma}_b \} = 2{\delta}_{ab}$. $J$ is the $4 \times 4$ symplectic matrix $J= -{\Gamma}_3 {\Gamma}_4$. An explicit realisation of these matrices is given in ~\cite{gravpre}.
\newpage
\underline{Parameters}
\para
There are two parameters we may introduce into our quantum mechanics which are of particular importance for us, the quintuplet of \emph{real masses} and the triplet of \emph{Fayet-Iliopoulos parameters}. We introduce these in turn.
\para
Firstly, we may separate the $D4$-branes, moving one to ${\vec{X}} = -{\vec{M}}$ and the other to ${\vec{X}} = {\vec{M}}$, so that their separation is $ 2 \pi {\alpha}^{\prime} (2M)$. In the quantum mechanics, this corresponds to weakly gauging the $SU(2)_F$ flavour symmetry, which results in the alteration to the Lagrangian
\be
\sum_{i=1}^2 {\vec{X}}^2 (|{\phi}_i|^2 +|{\tilde{\phi}}_i|^2) \rightarrow |{\vec{X}}-{\vec{M}}|^2 ( |{\phi}_1|^2 + |\tilde{\phi}_1|^2) + |{\vec{X}}+{\vec{M}}|^2 (  |{\phi}_2|^2 + |{\tilde{\phi}}_2|^2 )
\ee
for the bosons. The fermionic Lagrangian changes as;
\be
\sum_{i=1}^2 {\bar{\Psi}}_i ({\vec{X}} \cdot {\vec{\Gamma}}) {\Psi}_i \rightarrow {\bar{\Psi}}_1 (\vec{X} - {\vec{M}})\cdot {\vec{\Gamma}} {\Psi}_1 + {\bar{\Psi}}_2 (\vec{X}+{\vec{M}})\cdot {\vec{\Gamma}} {\Psi}_2.
\ee
\para
We may also turn on a background NS-NS $B$-field ${\cal{B}}_{\mu \nu}$ parallel to the $D4$-branes without breaking supersymmetry. In the quantum mechanics, this corresponds to including a triplet ${\bf r} = (r_1,r_2,r_3)$ of real Fayet-Iliopoulos parameters which obey ~\cite{Aharony:1997an, Lambert:1999ix}
\be
r_a = {\cal{B}}_{\mu \nu} {\eta}_a^{\mu \nu},
\ee
where ${\eta}_a^{\mu \nu}$ are the 't Hooft matrices (which can be found e.g. in Appendix A of ~\cite{gravpre}).
When $r_a = 0$, we may write the quartic terms for ${\phi}_i$ and ${\tilde{\phi}}_i$ in the action (\ref{hyper}) in a manifestly $SU(2)_R$-covariant way as
\be
-\frac{g^2}{2} \sum_{i=1}^2|{\omega}_i {\tau}^a {\omega}_i|^2
\ee
where ${\omega}_i = ({\phi}_i, {\tilde{\phi}}_i^{\dagger} )^T$ transforms as a doublet of $SU(2)_R$. Turning on the background $B$-field changes this term to
\be
-\frac{g^2}{2} \sum_{i=1}^2|{\omega}_i {\tau}^a {\omega}_i|^2 &\rightarrow& -\frac{g^2}{2} \sum_{i=1}^2|{\omega}_i {\tau}^a {\omega}_i - r^a |^2 ,
\ee
Explicitly, in terms of the component fields this becomes
\be
-\frac{g^2}{2} \left( \sum_{i=1}^2 |{\phi}_i|^2 + |{\tilde{\phi}}_i|^2 - r^3 \right) ^2 - 2g^2 \left| \sum_{i=1}^2 {\phi}_i {\tilde{\phi}}_i - (r^1+ir^2) \right| ^2 .
\ee

\vspace{2ex}

\subsection{Ground States}

Our model has a Bosonic potential;
\be
V &=& |{\vec{X}}-{\vec{M}}|^2 (|{\phi}_1|^2 + |{\tilde{\phi}}_1|^2) + |{\vec{X}}+{\vec{M}}|^2 ( |{\phi}_2|^2 + |{\tilde{\phi}}_2|^2) \\ &\:& \;\;\; + \frac{g^2}{2} (\sum_{i=1}^2 |{\phi}_i|^2 - |{\tilde{\phi}}_i|^2 -r)^2 + 2g^2|\sum_{i=1}^N {\tilde{\phi}}_i {\phi}_i|^2 .
\label{bosac}
\ee
We have switched on only the $3$-component of the Fayet-Iliopoulos parameters. We can see that there are two distinct classical vacua, both of which have ${\tilde{\phi}}_i = 0$
\be
\text{Vacuum 1} \;\;\; |{\phi}_1|^2 = r \;\; , \;\; |{\phi}_2|^2 = 0 \,\, ; \,\, {\vec{X}}=+{\vec{M}}, \nonumber \\
\text{Vacuum 2} \;\;\; |{\phi}_1|^2 = 0 \;\; , \;\; |{\phi}_2|^2 = r \,\, ; \,\, {\vec{X}}=-{\vec{M}},
\ee
which corresponds to the $D0$-brane living inside one or other of the $D4$-branes. The Witten Index ensures that both these states survive in the quantum theory. There can be no tunneling between these two vacua because the kink solution carries fermionic zero-modes which must be saturated in the path integral in order for us to get a non-zero overlap. We investigate these zero-modes in more detail in section 3. We now study these two separated vacua, and the spectrum of excited states living above them.

\vspace{2ex}

\subsection{The Spectrum}
\para
Our system has a dimensionless coupling $1/|{\vec{M}}|r$, such that when $|{\vec{M}}|r >>1$, the physics around each vacuum is described by a single, free hypermultiplet. This corresponds to the fact that when the $D0$-brane lives inside a $D4$-brane, the $D0$-$D4$-string states between the two branes have a mass $g^2 r$, and condense out of the spectrum. We may understand this quantum mechanically as the associated degrees of freedom being eaten by the Higgs mechanism, and giving a mass to the remaining modes in the opposite hypermultiplet.
\para
\underline{Vacuum 1}
\para
Explicitly, in Vacuum 1, ${\phi}_1$ is eaten by the Higgs mechanism, and the remaining dynamical degree of freedom is ${\Phi}_2$ with mass $2{\vec{M}}$. In this vacuum, the fermionic Hamiltonian is
\be
H_F^1 = 2 {\bar{\Psi}}_2 ({\vec{M}} \cdot {\vec{\Gamma}}) {\Psi}_2,
\label{fham1}
\ee
and we can use this, along with the canonical bracket for fermions $\{ {\Psi}_{2 \alpha}, {\bar{\Psi}}_{2 \beta} \} = {\delta}_{\alpha \beta}$ to build the usual tower of states. We define the reference state $|0 \rangle$ to obey ${\Psi}_{1 \alpha} |0 \rangle = {\Psi}_{2 \alpha} |0 \rangle = 0$, so that our tower of states is;
\para
\begin{center}
\begin{tabular}{||c|c|c||} \hline
State & Multiplicity & Eigenvalue \\ \hline
$|0\rangle$ & $1$ & $0$ \\ \hline
${\bar{\Psi}}_{2 \alpha}|0\rangle$ & $4$ & $(-2|{\vec{M}}|)_2, (+2|{\vec{M}}|)_2$ \\ \hline
${\bar{\Psi}}_{2 \alpha}{\bar{\Psi}}_{2 \beta}|0\rangle$ & $6$ & $-4|{\vec{M}}|, 0_4, +4|{\vec{M}}|$ \\ \hline
${\bar{\Psi}}_{2 \alpha}{\bar{\Psi}}_{2 \beta}{\bar{\Psi}}_{2 \gamma}|0\rangle$ & $4$ & $(-2|{\vec{M}}|)_2, (+2|{\vec{M}}|)_2$ \\ \hline
${\bar{\Psi}}_{2 \alpha}{\bar{\Psi}}_{2 \beta}{\bar{\Psi}}_{2 \gamma}{\bar{\Psi}}_{2 \delta}|0\rangle$ & $1$ & $0$ \\ \hline
\end{tabular}
\end{center}
\vspace{2ex}

\para
The lowest-lying state in the spectrum, the susy vacuum $|{\Omega}_1 \rangle$, lives in the middle sector ${\bar{\Psi}}_{2 \alpha} {\bar{\Psi}}_{2 \beta} |0 \rangle$ and has eigenvalue $-4|{\vec{M}}|$ under $H_F^1$. To this we must add the zero-point energy of the two complex bosons ${\phi}_2$ and ${\tilde{\phi}}_2$ in the spectrum. Each of these bosons contributes $+2M$, so that the state $|{\Omega}_1 \rangle$ has zero energy, as we would expect of the susy vacuum. The explicit fermi structure of $|{\Omega}_1 \rangle$ depends on ${\vec{M}}$ and is given in Appendix A.
\para
The main focus of our discussion will be the first-excited states of the theory. There are four such states, which divide naturally into two pairs, one pair living in the $1$-fermi ${\bar{\Psi}}_{2 \alpha} |0 \rangle$-sector, and the other pair in the $3$-fermi ${\bar{\Psi}}_{2 \alpha} {\bar{\Psi}}_{2 \beta} {\bar{\Psi}}_{2 \gamma}|0 \rangle$-sector. These states all have eigenvalue $-2|{\vec{M}}|$ under $H_F^1$, so that with the contribution of the bosonic zero-point energy, they have total energy $+2M$. The explicit fermi structure of the first excited states is also given in Appendix A. Schematically, we may write them as
\be
| 1 \rangle = {\Psi}_{2 \alpha} |{\Omega}_1 \rangle \;\;\; \text{and} \;\;\; | ^{\star} 1 \rangle = {\bar{\Psi}}_{2 \alpha} |{\Omega}_1 \rangle
\ee

\para
\underline{Vacuum 2}
\para
In Vacuum 2, the roles are reversed, and it is ${\phi}_2$ that is eaten by the Higgs mechanism, leaving ${\Phi}_1$ as the dynamical hypermultiplet with mass $-2{\vec{M}}$. Our fermionic Hamiltonian is this time
\be
H_F^2 = -2{\bar{\Psi}}_1 ( {\vec{M}} \cdot {\vec{\Gamma}}) {\Psi}_1 ,
\ee
The lowest-lying state in this sector is $|{\Omega}_2 \rangle$ which has eigenvalue $-4|{\vec{M}}|$ under $H_F^2$, and lives in ${\bar{\Psi}}_{1 \alpha}{\bar{\Psi}}_{1 \beta}|0 \rangle$. When dressed with the bosons in their ground state, this state again has zero energy. There are four first excited states, two in the $1$-fermi ${\bar{\Psi}}_{1 \alpha}|0 \rangle$-sector and two in the $3$-fermi
${\bar{\Psi}}_{1 \alpha}{\bar{\Psi}}_{1 \beta}{\bar{\Psi}}_{1 \gamma}|0 \rangle$-sector. The explicit fermi structure of these states is again given in Appendix A. Schematically we denote them as
\be
| 2 \rangle = {\Psi}_{1 \alpha} |{\Omega}_2 \rangle \;\;\; \text{and} \;\;\; | ^{\star} 2 \rangle = {\bar{\Psi}}_{1 \alpha} |{\Omega}_2 \rangle
\ee
\para
We show in section 3 that the BPS kinks in our quantum mechanics carry a total of four fermi zero-modes. In order to saturate the zero-mode contribution to the instanton path-integral, we need to sandwich four fermions between the two vacua. The overlap of the ground states is $\langle {\Omega}_1 | {\Omega}_2 \rangle$, and so it leaves all the Grassmann integrals unsaturated, and receives no BPS kink correction. Similarly, the overlap of the first excited states is schematically $\langle {\Omega}_1 |{\bar{\Psi}}_{1 \alpha} {\Psi}_{2, \beta} | {\Omega}_2 \rangle$, which only saturates two Grassmann integrals, and again cannot receive any correction.
\para
It is possible that the kinks could correct the second-excited states in the spectrum, since their overlaps would carry a total of four fermions. However we will show that lower-lying states in the spectrum may receive BPS corrections if we allow the $D4$-branes to move.
\vspace{2ex}
\section{Berry's Phase}
\para
We now turn to the possibility of adiabatically moving the $D4$-branes around a closed path in the transverse space. Under such an evolution, we expect that the states of a given energy $\{ |a_i \rangle \}$ will mix with one another, undergoing a Berry holonomy
\be
| a_i \rangle \rightarrow P \exp \left( -i \oint ({\omega}_{\mu})_{ij}({\vec{M}}) \, dM^{\mu} \right) |a_j \rangle
\ee
given in terms of a non-Abelian Berry connection ${\omega}_{\mu}({\vec{M}})$.

\subsection{Perturbative Berry Phase}
We first study the possible perturbative Berry phase the $D4$-$D0$-$D4$ system can admit.
Under adiabatic motion of the $D4$-brane in the semiclassical limit $Mr >>1$, where the vacua are far-separated, it can be shown that the vacuum states of the system have no Abelian Berry phase ~\cite{gravpre}. It is the first-excited states that undergo a non-Abelian holonomy. We will focus solely on the $1$-fermion states living above \emph{both} vacua \footnote{It can be shown that these $1$-fermi states do not mix with the first-excited states in the $3$-fermi sector. We may in fact derive identical results for the Berry phase, both perturbative and non-perturbative, in the $3$-fermi sector.}. We first look at the pair of first-excited $1$-fermi states above vacuum 1, where we find that the fermionic Hamiltonian (\ref{fham1}) acts as
\be
H_F^1 |_{{\bar{\Psi}}_{2 \alpha}|0 \rangle} = 2{\vec{M}} \cdot {\vec{\Gamma}}.
\ee
We define projectors onto the orthogonal eigenspaces of eigenvalue $\pm 2|{\vec{M}}|$ under this Hamiltonian, they are
\be
P_{\pm} = 1 \pm \frac{{\vec{M}} \cdot {\vec{\Gamma}}}{|{\vec{M}}|}.
\ee
Our two lowest-lying states in this sector are then $|a\rangle = P_- {\bar{\Psi}}_{2,1}|0 \rangle$ and $|b\rangle = P_- {\bar{\Psi}}_{2,3}|0 \rangle$, and are valid everywhere except along ${\vec{M}} = (0,0,0,0,-M)$ where the lowest-lying states are orthogonal to both ${\bar{\Psi}}_{2,1}|0\rangle$ and ${\bar{\Psi}}_{2,3}|0\rangle$. Because these states carry the projector $P_-$, it can be shown that the relevant non-Abelian connection is that of the \emph{Yang monopole} (see ~\cite{gravpre,Chen:1998qb});
\be
(A_{\mu})_{ab} = \frac{-M^{\nu}}{2M(M+M_5)} {\eta}^m_{\mu \nu} {\tau}^m_{ab},
\ee
where ${\eta}^m_{\mu \nu}$ are the self-dual 't Hooft matrices, and ${\tau}^m_{ab}$ are the usual Pauli matrices.
\para
Above Vacuum 2, we have another pair of first-excited $1$-fermion states. In this sector, the fermionic Hamiltonian $H_F^2$ acts as
\be
H_F^2 |_{{\bar{\Psi}}_{1 \alpha}|0 \rangle} = -2{\vec{M}} \cdot {\vec{\Gamma}}.
\ee
and the lowest-lying states are now $|c\rangle = P_+ {\bar{\Psi}}_{1,2}|0 \rangle$ and $|d\rangle = P_+ {\bar{\Psi}}_{1,4}|0 \rangle$. These states again go bad when ${\vec{M}} = (0,0,0,0,-M)$. Since they carry a projector $P_+$, it can be shown that for these states, the connection is that of the \emph{anti-Yang} or \emph{Yin} monopole
\be
({\tilde{A}}_{\mu})_{ab} = \frac{-M^{\nu}}{2M(M+M_5)} {\bar{\eta}}^m_{\mu \nu} {\tau}^m_{ab} = (A_{\mu}^{\star})_{ab},
\ee
where ${\bar{\eta}}^m_{\mu \nu}$ are the anti-self-dual 't Hooft matrices. The total connection in this sector is then given by
\be
({\omega}_{\mu})_{ab} = \left( \begin{array}{cc} \langle 2 | {\partial}_{\mu}| 2 \rangle & \langle 2 | {\partial}_{\mu}| 1 \rangle \\ \langle 1 | {\partial}_{\mu}| 2 \rangle & \langle 1 | {\partial}_{\mu}| 1 \rangle \end{array} \right) = \left( \begin{array}{cc} {\tilde{A}}_{\mu} & 0 \\ 0 & A_{\mu} \end{array} \right),
\label{berryconn}
\ee
where we have used the abbreviations $|1\rangle = \{ |a \rangle, |b \rangle \}$ to mean the pair of one-fermi states above Vacuum 1, and $|2 \rangle = \{ |c \rangle, |d \rangle \}$ to mean the two one-fermi states above Vacuum 2.
\newpage
\underline{Singular Gauge Transformation}
\para
This connection for the first excited $1$-fermi states above the two vacua may be rotated to a spherically symmetric gauge by means of the gauge transformation ~\cite{smallzhang}
\be
U = V \exp \biggl( \frac {i \theta M^{\nu}{\Gamma}_{\nu 5}}{\sqrt{M^2-M_5^2}} \biggr) \;\;\; , \;\;\; V =\left( \begin{array}{cccc} 1 & 0 & 0 & 0 \\ 0 & 0 & 1 & 0 \\ 0 & 1 & 0 & 0 \\ 0 & 0 & 0 & 1 \end{array} \right),
\label{ugauge}
\ee
where $\cos (\theta ) = M_5/M$. Under $U$, ${\omega}_{\mu}$ transforms as
\be
{\omega}_{\mu} \rightarrow U {\omega}_{\mu} U^{\dagger} - i({\partial}_{\mu} U) U^{\dagger} = \frac{M^{\nu}{\Gamma}_{\mu \nu}}{X^2},
\ee
where ${\Gamma}_{\mu \nu} = \frac{1}{4i} [ {\Gamma}_{\mu}, {\Gamma}_{\nu} ]$ are the Lorentz generators of the group $SO(5)$.

\subsection{Non-Perturbative Corrections}
\para
In a previous paper ~\cite{geophase}, it was shown that the Berry phase connection for the ${\mathbb{CP}}^1$ sigma model was corrected by BPS instantons. We also computed the one-instanton correction, and explained its role in desingularising the Dirac monopole connection we found at the perturbative level. We perform a similar computation for the first excited states in our current system below. We first study the classical BPS kink configuration.
\para
\underline{Bogomolnyi Equations}
\para
We take the bosonic action (\ref{bosac}) and choose all excitations of ${\vec{X}}$ to live solely in the direction ${\vec{X}} = (0,0,0,0,X)$. We will ignore the fields ${\tilde{\phi}}_i$ which play no dynamical role in the kink solution. This gives the Euclidean action
\be
S =  \int \, dt \left( \frac{1}{2g^2} {\dot{X}}^2 + \sum_{i=1}^2 |{\mathcal{D}}_t{\phi}_i|^2 + |X-M_i|^2 |{\phi}_i|^2 + \frac{g^2}{2} \biggl( \sum_{i=1}^2 |{\phi}_i|^2 - r {\biggr)}^2 \right).
\ee
By completing the square, we obtain
\be
S = \int \, dt \left( \frac{1}{2g^2} \left( \dot{X} \mp g^2 ( \sum_{i=1}^2 |{\phi}_i|^2 - r ) \right)^2 +\left| \sum_{i=1}^2 {\mathcal{D}}_t {\phi}_i \mp (X-M_i){\phi}_i \right| ^2 \right) +T ,
\ee
where  $T = 2Mr$  is a topological charge. By saturating the bound on the action that S=T, we find the Bogomolnyi equations
\be
{\dot{X}} = \pm g^2 \biggl(\sum_{i=1}^2 |{\phi}_i|^2 -r {\biggr)}^2 \;\;\; , \;\;\; {\mathcal{D}}_t {\phi}_i = \pm (X-M_i){\phi}_i .
\label{bogeqn}
\ee
and the action of the kink is $S_{kink} = T$. We choose the upper signs in the Bogomolnyi equations, corresponding to the kink solution rather than the anti-kink. This solution interpolates between Vacuum 1 at $t=-\infty$ and Vacuum 2 at $t=+\infty$.
\para
\underline{Fermion Zero Modes}
\para
Fundamental to instanton computations are the fermion zero-modes, which tell us which quantities can receive corrections from BPS instantons. We therefore analyse the fermionic modes of our system. Because of the enhanced supersymmetry of this system each bosonic zero mode will give us two fermionic zero-modes, and so in order to soak up all the Grassmann integrations in the kink measure, we need to make a four-fermion insertion.
\para
We therefore need to know which of the fermion Dirac operators can have zero-modes. The Yukawa terms govern how the
fermions interact, and so the forms of the Dirac operators. In the background where ${\tilde{\phi}}_i = 0$, they
are
\be
L_{Yukawa} &= i \sqrt{2} {\phi}_i ({\bar{\lambda}}_-{\bar{\psi}}_{i+} - {\bar{\lambda}}_+{\bar{\psi}}_{i-})
-\sqrt{2}{\phi}_i({\eta}_-{\tilde{\psi}}_{i+} - {\eta}_+{\tilde{\psi}}_{i-}) \nonumber \\ &\;\;\;\;\; + i\sqrt{2}{\phi}_i^{\dagger}({\psi}_{i-}{\lambda}_+ - {\psi}_{i+}{\lambda}_-) - \sqrt{2}{\phi}_i^{\dagger}({\bar{\tilde{\psi}}}_{i+}{\bar{\eta}}_- - {\bar{\tilde{\psi}}}_{i-}{\bar{\eta}}_+).
\ee
Including the usual kinetic terms for the fermions, we find that there are four Dirac equations;
\be \Delta \left(\begin{array}{c}\lambda_-
\\ \bar{\psi}_{i+}\end{array}\right) =
\Delta^\dagger\left(\begin{array}{c}\lambda_+
\\ \bar{\psi}_{i-}\end{array}\right)  =
\Delta \left(\begin{array}{c} i\eta_+
\\ \bar{\tilde{\psi}}_{i-}\end{array}\right) =
\Delta^\dagger\left(\begin{array}{c}-i\eta_-
\\ \bar{\tilde{\psi}}_{i+}\end{array}\right)  = 0,
\ee
where
\be \Delta =
 \left(\begin{array}{cc} \frac{1}{g^2}\,\partial_\tau
&
 -i\sqrt{2} \phi_i \\ i\sqrt{2}\bar{\phi}_i\ \ \ \  & -{\cal D}_\tau
 + (X-M_i) \end{array}\right),\ \
\Delta^\dagger = \left(\begin{array}{cc}
-\frac{1}{g^2}\,\partial_\tau &
 -i\sqrt{2} \phi_i \\ i\sqrt{2}\bar{\phi}_i\ \ \ \  & {\cal D}_\tau
 +
 (X-M_i) \end{array}\right).\ \ \ \ \label{dops}\ee
\para
In the first pair of equations, our $(2,2)$ analysis ~\cite{geophase} shows that ${\Delta}$ has zero modes whereas ${\Delta}^{\dagger}$ does not, so that the pairs of fermions $({\lambda}_-,{\bar{\psi}}_{i+})$ and $({\bar{\lambda}}_+,{\psi}_{i-})$ carry the fermionic zero modes. The pairs $({\eta}_+, {\bar{\tilde{\psi}}}_{i-})$ and $({\bar{\eta}}_-, {\tilde{\psi}}_{i+})$ thus also carry zero-modes. For finite $g^2$ our zero modes are
\be
{\xi}_- &= \frac{1}{g^2} {\lambda}_- + {\bar{\psi}}_{i+} \;\;\; , \;\;\; {\bar{\xi}}_+ = \frac{1}{g^2} {\bar{\lambda}}_+ + {\psi}_{i-}, \nonumber \\
{\tilde{\xi}}_+ &= \frac{1}{g^2} {\eta}_+ + {\bar{\tilde{\psi}}}_{i-} \;\;\; , \;\;\; {\bar{\tilde{\xi}}}_- =
\frac{1}{g^2} {\bar{\eta}}_- + {\tilde{\psi}}_{i+}.
\ee
In the $g^2 \rightarrow \infty$ limit, the correlator that can get instanton corrections is therefore
\be
\langle {\psi}_{-}{\bar{\psi}}_+ {\tilde{\psi}}_+ {\bar{\tilde{\psi}}}_- \rangle .
\label{correl}
\ee
\para
\underline{Berry's Phase}
\para
We now wish to calculate the Berry phase for the four $1$-fermi first-excited states in this non-perturbative regime. The Berry phase connection for these four states $|a \rangle = \{ |1 \rangle, | 2 \rangle \}$ is given by
\be
({\omega}_{\mu})_{ab} = \langle a | \frac{\partial}{\partial M^{\mu}}| b \rangle
\ee
In differentiating with respect to ${\vec{M}}$ we bring down two fermi zero modes for the system.
We first calculate the Berry phase overlaps for the ground states, and see that the overlap contains two-few fermions to get corrections from the BPS kinks. Hence there is no correction to the vacuum Berry phase under adiabatic motion of the $D4$-brane, and so the Berry phase connection for these states remains zero to all orders.
\para
The first-excited states can get correction from (\ref{correl}), since each excited state carries a fermi zero mode, and we get an extra two from the differentiation, making a total of four fermi zero modes in the overlap. In order to simplify our computations, we choose to set $M_5 = M$ (note that this is ok as the Dirac string lies along $M_5 = -M$), which has the added benefit that the perturbative contribution to the Berry connection vanishes with this choice. The explicit calculation of the overlaps is relegated to Appendix A; the result is that the one-instanton Berry phase connection is
\be
({\omega}_1)_{ij}&=&f {\tau}^1 \otimes {\tau}^2/2M , \nonumber \\
({\omega}_2)_{ij}&=&f {\tau}^2 \otimes {\tau}^0/2M , \nonumber \\
({\omega}_3)_{ij}&=&f {\tau}^1 \otimes {\tau}^3/2M , \nonumber \\
({\omega}_4)_{ij}&=&f {\tau}^1 \otimes {\tau}^1/2M , \nonumber \\
({\omega}_5)_{ij}&=&0
\ee
\para
The function $f=f(Mr) = K \langle {\epsilon}_- {\bar{\epsilon}}_+ {\tilde{\epsilon}}_+ {\bar{\tilde{\epsilon}}}_- \rangle$ is determined by an explicit one-loop instanton calculation. The exact functional form of $f(Mr)$ is computed in Appendix B, and gives the result that
\be
f(Mr) = \frac{3 \pi M r}{16} e^{-2Mr}
\ee
\para
However, notice that we have performed this calculation in \emph{singular gauge}. By rotating our axes so that ${\vec{M}} = (0,0,0,0,M)$, we have effectively ignored the physics of the system in the direction parallel to $M_5$. In order that we capture the instanton corrections to all of the physical quantities, we must now transform back to the spherically symmetric configuration. We achieve this by means of the gauge transformation (\ref{ugauge}) under which the non-perturbative part ${\Sigma}_{\mu}$ of our connection becomes (on setting $\vec{M} = (0,0,0,0,M)$)
\be
{\Sigma}_{\mu} \rightarrow U{\Sigma}_{\mu} U^{\dagger} = -\frac{f}{M} {\Gamma}_{\mu 5}.
\ee
Note that the usual derivative part of the gauge transformation has already been used in transforming the perturbative connection. Including the perturbative effects, the full connection up to one-instanton is given by
\be
{\omega}_{\mu} = [1-f(Mr)]\frac{M^{\nu} {\Gamma}_{\mu \nu}}{M^2}
\label{tchrakian}
\ee
which is the connection for the \emph{Tchrakian} $SO(5)$ monopole ~\cite{Tchrakian:2006nn}. It is important to note that, whereas previously we had the freedom to write the connection in any gauge we wished, we are now forced to use the $SO(5)$-covariant gauge in order that we capture all of the physics of the instanton corrections.
\para

\vspace{2ex}

\section{Discussion}

The appearance of the Lorentz generators ${\Gamma}_{\mu \nu}$ of $SO(5)$ in the one-instanton connection signals an interesting development; an emergent $SO(5)$ structure from a calculation that, at least \emph{a priori}, we would have expected to yield a $U(4)$ connection. This has some deep connection to the isometry group of the transverse ${\mathbb{R}}^5$, but the appearance of this $SO(5)$ structure is not well-understood. It would be very interesting to know if this $SO(5)$ emergent geometry persists in the all-instanton connection.
\para
There does exist a natural $SO(5)$-connection over ${\mathbb{R}}^5$, which has been described by  ~\cite{Kihara:2004yz}. It shares its asymptotic behaviour with our connection, but is described in full by the Bogomolnyi equation
\be
{\star}_5 (F \wedge F) = {\cal{D}} {\xi} ,
\ee
for some scalar field $\xi$. Spherically symmetric solutions of this equation take the form (\ref{tchrakian}), except that $f \sim e^{-r^3 M^3 /3}$ for large $M$. We cannot generate such a fall-off from our instanton effects at any order, since they are always weighted by the instanton action $e^{-2Mr}$, and so our solution is not the BPS object in $SO(5)$ gauge theory.
\para
The effects of the instanton correction are somewhat mysterious. In the 't Hooft-Polyakov case, the one-instanton correction was the leading order effect that, when calculated to all orders would have guaranteed the smoothness of the connection at the origin. Indeed, this calculation has now been done, and the result to all-instanton level is the BPS 't Hooft-Polyakov monopole \cite{Sonner:2008be}. In our current case, things are somewhat more blurred. The first-excited states considered as wavepackets have a width $\sigma \sim 1/M$. As we take $M \rightarrow 0$, these states become non-normalisable since their compact support near the origin spreads to infinity in field space, and strictly when $M=0$ they leave the physical Hilbert space of our theory altogether. Since we are calculating a Berry phase connection for these states, we should perhaps expect the connection to become singular, signaling that it knows about this bad behaviour of the first-excited states. It would be nice to see what effect the all-instanton correction would have in the $D4$-$D0$-$D4$ case.
\vfill

\newpage
\section*{Appendix A:  Vacua and Excited States}

\setcounter{section}{1}
\renewcommand{\theequation}{\Alph{section}.\arabic{equation}}

In this appendix, we study the two disjoint Hilbert spaces generated by the fermionic creation operators ${\bar{\Psi}}_1$ and ${\bar{\Psi}}_2$ separately.
\para
\underline {Vacuum 1; $\bar{\Psi}_2$}
\para
The supersymmetric vacuum is the state living the middle sector ${\bar{\Psi}}_{2 \alpha} {\bar{\Psi}}_{2 \beta} |0 \rangle$, and has eigenvalue $-4M$ under $H_{F1}$. It takes the explicit form
\be
| {\Omega}_1 \rangle &= \frac{1}{2M} [(M_1 - iM_4) {\bar{\Psi}}_{2,1} {\bar{\Psi}}_{2,2} + (M_5-M) {\bar{\Psi}}_{2,1} {\bar{\Psi}}_{2,3} +(M_2 +iM_3) {\bar{\Psi}}_{2,1} {\bar{\Psi}}_{2,4} \nonumber \\
&\;\;\; +(M_2-iM_3) {\bar{\Psi}}_{2,2} {\bar{\Psi}}_{2,3} + (-M_5-M) {\bar{\Psi}}_{2,2} {\bar{\Psi}}_{2,4} + (M_1+iM_4) {\bar{\Psi}}_{2,3} {\bar{\Psi}}_{2,4} ] |0 \rangle .
\ee
The states of interest are the four first-excited states, two of which live in the ${\bar{\Psi}}_{2 \alpha} |0 \rangle \}$-sector, where $H_F^1$ acts as $-2{\vec{M}} \cdot {\vec{\Gamma}}$. These take the form
\be
P_- {\bar{\Psi}}_{2,2} | 0 \rangle &= \frac{1}{N} [ (-M_2 +iM_3) {\bar{\Psi}}_{2,2} +(M+M_5) {\bar{\Psi}}_{2,2} + (-M_1 + iM_4) {\bar{\Psi}}_{2,3} ]|0 \rangle , \nonumber \\
P_- {\bar{\Psi}}_{2,4} | 0 \rangle &= \frac{1}{N} [ (M_1 + iM_4) {\bar{\Psi}}_{2,1} + (-M_2 -iM_3) {\bar{\Psi}}_{2,3} + (M+M_5) {\bar{\Psi}}_{2,4} ] |0 \rangle,
\ee
The other pair of states live in the $ ^{\star}{\bar{\Psi}}_{2,\alpha}|0 \rangle = J_{\alpha}^{\beta} {\epsilon}_{\beta \gamma \delta \rho}{\bar{\Psi}}_{2,\gamma} {\bar{\Psi}}_{2,\delta} {\bar{\Psi}}_{2,\rho} |0 \rangle$ - sector, where $H_F^1$ acts as $-2{\vec{M}} \cdot {\vec{\Gamma}}$. These two states are
\be
P_+ ( ^{\star} {\bar{\Psi}}_{2,1} ) |0\rangle &= \frac{1}{N} [ (M+M_5){\bar{\Psi}}_{2,4} {\bar{\Psi}}_{2,1} {\bar{\Psi}}_{2,2} + (M_2+iM_3) {\bar{\Psi}}_{2,1} {\bar{\Psi}}_{2,2} {\bar{\Psi}}_{2,3}, \nonumber \\
& \;\;\; - (-M_1 + iM_4) {\bar{\Psi}}_{2,3} {\bar{\Psi}}_{2,4} {\bar{\Psi}}_{2,1} ] |0 \rangle \nonumber \\
P_+ ( ^{\star} {\bar{\Psi}}_{2,3} ) |0\rangle &= \frac{1}{N} [ (M_1 + iM_4) {\bar{\Psi}}_{2,1} {\bar{\Psi}}_{2,2} {\bar{\Psi}}_{2,3} - (M+M_5) {\bar{\Psi}}_{2,1} {\bar{\Psi}}_{2,2} {\bar{\Psi}}_{2,3} \nonumber \\
& \;\;\; -(M_2 -iM_3) {\bar{\Psi}}_{2,3} {\bar{\Psi}}_{2,4} {\bar{\Psi}}_{2,1} ] |0 \rangle,
\ee
Note that all these states have been normalised, with normalisation factor \newline
$N = {\sqrt{2M(M+M_5)}}$ so that they have Dirac strings lying along the $M_5 = -M$ axis.
\para
\underline{Vacuum 2; ${\bar{\Psi}}_{1 \alpha}$}
\para
Again, we also need to know the lowest-lying state in this Hilbert space, it is again in the middle sector ${\bar{\Psi}}_{1 \alpha} {\bar{\Psi}}_{1 \beta}$ with eigenvalue $-4M$ under $H_{F2}$, and is given explicitly by
\be
|{\Omega}_2 \rangle &= \frac{1}{2M} [ (M_1 + iM_4) {\bar{\Psi}}_{1,1} {\bar{\Psi}}_{1,2} + (M+M_5) {\bar{\Psi}}_{1,1} {\bar{\Psi}}_{1,3} + (M_2 - iM_3) {\bar{\Psi}}_{1,1} {\bar{\Psi}}_{1,4} \nonumber \\
& \;\;\;+ (M_2 + iM_3) {\bar{\Psi}}_{1,2} {\bar{\Psi}}_{1,3} + (M-M_5) {\bar{\Psi}}_{1,2} {\bar{\Psi}}_{1,4} + (M_1 - iM_4) {\bar{\Psi}}_{1,3} {\bar{\Psi}}_{1,4} ]|0 \rangle.
\ee
The 1-fermi first excited states above Vacuum 2 live in the ${\bar{\Psi}}_{1, \alpha} |0 \rangle$-sector, where $H_F^2$ acts as $-2{\vec{M}} \cdot {\vec{\Gamma}}$, and are given explicitly (when normalised to $1$) by
\be
P_+ {\bar{\Psi}}_{1,1} |0 \rangle &= \frac{1}{N} [(M+M_5){\bar{\Psi}}_{1,1} + (M_2 + i M_3){\bar{\Psi}}_{1,2} + (-M_1+ i M_4){\bar{\Psi}}_{1,4}]|0 \rangle, \nonumber \\
P_+ {\bar{\Psi}}_{1,3} |0 \rangle &=  \frac{1}{N} [(M_1 + iM_4){\bar{\Psi}}_{1,2} + (M+M_5){\bar{\Psi}}_{1,3} + (M_2-iM_3){\bar{\Psi}}_{1,4}]|0 \rangle.
\label{vac11}
\ee
The other two states live in the three-fermi sector $ ^{\star}{\bar{\Psi}}_{1,\alpha}|0 \rangle = J_{\alpha}^{\beta} {\epsilon}_{\beta \gamma \delta \rho}{\bar{\Psi}}_{1,\gamma} {\bar{\Psi}}_{1,\delta} {\bar{\Psi}}_{1,\rho} |0 \rangle$-sector, where $H_F^2$ acts as $2{\vec{M}} \cdot {\vec{\Gamma}}$. These two states are
\be
P_- (^{\star}{\bar{\Psi}}_{1,2})|0 \rangle &= \frac{1}{N} [ (-M_2 + i M_3){\bar{\Psi}}_{1,4}{\bar{\Psi}}_{1,1}{\bar{\Psi}}_{1,2}+(M+M_5){\bar{\Psi}}_{1,1}{\bar{\Psi}}_{1,2}{\bar{\Psi}}_{1,3} \nonumber \\ & \;\;\; +(-M_1 + iM_4){\bar{\Psi}}_{1,2}{\bar{\Psi}}_{1,3}{\bar{\Psi}}_{1,4}] |0\rangle, \nonumber \\
P_- ( ^{\star}{\bar{\Psi}}_{1,4})|0 \rangle &= \frac{1}{N}[ (M_1+iM_4){\bar{\Psi}}_{1,4}{\bar{\Psi}_{1,1}{\bar{\Psi}}_{1,2} + (M_2 + i M_3){\bar{\Psi}}_{1,2}{\bar{\Psi}}_{1,3} {\bar{\Psi}}}_{1,4} \nonumber \\
& \;\;\; - (M+M_5){\bar{\Psi}}_{1,3}{\bar{\Psi}}_{1,4}{\bar{\Psi}}_{1,1} ] | 0 \rangle,
\label{vac12}
\ee
\para
We now need to differentiate these first excited states in order that we can see how the instanton contributions can correct the Berry phase connection
\be
(A_{\mu})_{ab} = i \langle a| \frac{\partial}{\partial M^{\mu}} | b \rangle ,
\ee
The derivatives of the states in Vacuum 1, on setting $M_5 = M$, are
\be
\frac{\partial}{\partial M^1} P_- {\bar{\Psi}}_{2,2} |0 \rangle &= \frac{1}{2M} [ {\bar{\Psi}}_{2,3} {\Psi}_{2,4} {\Psi}_{2,2} + {\Psi}_{2,1} + {\bar{\Psi}}_{2,2} {\Psi}_{2,4} {\Psi}_{2,3} ] |{\Omega}_1 \rangle, \nonumber \\
\frac{\partial}{\partial M^1} P_- {\bar{\Psi}}_{2,4} |0 \rangle &= \frac{1}{2M} [ -{\bar{\Psi}}_{2,1} {\Psi}_{2,4} {\Psi}_{2,2} + {\bar{\Psi}}_{2,4} {\Psi}_{2,2} {\Psi}_{2,1} + {\Psi}_{2,3} ]|{\Omega}_1 \rangle, \nonumber \\
\frac{\partial}{\partial M^1} P_+ ( ^{\star} {\bar{\Psi}}_{2,1} |0 \rangle) &= \frac{1}{2M} [ {\bar{\Psi}}_{2,3} {\bar{\Psi}}_{2,1} {\Psi}_{2,2} + {\bar{\Psi}}_{2,4} + {\bar{\Psi}}_{2,1} {\bar{\Psi}}_{2,2} {\Psi}_{2,3} ] |{\Omega}_1 \rangle, \nonumber \\
\frac{\partial}{\partial M^1} P_+ ( ^{\star} {\bar{\Psi}}_{2,3} |0 \rangle) &= \frac{1}{2M} [ -{\bar{\Psi}}_{2,1} {\bar{\Psi}}_{2,3} {\Psi}_{2,4} + {\bar{\Psi}}_{2,3} {\bar{\Psi}}_{2,4} {\Psi}_{2,1} - {\bar{\Psi}}_{2,2} ] |{\Omega}_1 \rangle,
\ee
\be
\frac{\partial}{\partial M^2} P_- {\bar{\Psi}}_{2,2} |0 \rangle &= \frac{1}{2M} [ {\bar{\Psi}}_{2,1} {\Psi}_{2,4} {\Psi}_{2,2} + {\bar{\Psi}}_{2,2} {\Psi}_{2,4} {\Psi}_{2,1} - {\Psi}_{2,3} ] |{\Omega}_1 \rangle, \nonumber \\
\frac{\partial}{\partial M^2} P_- {\bar{\Psi}}_{2,4} |0 \rangle &= \frac{1}{2M} [ {\bar{\Psi}}_{2,3} {\Psi}_{2,4} {\Psi}_{2,2} + {\Psi}_{2,1} + {\bar{\Psi}}_{2,4} {\Psi}_{2,3} {\Psi}_{2,2}]|{\Omega}_1 \rangle, \nonumber \\
\frac{\partial}{\partial M^2} P_+ ( ^{\star} {\bar{\Psi}}_{2,1} |0 \rangle) &= \frac{1}{2M} [ -{\bar{\Psi}}_{2,1} {\bar{\Psi}}_{2,3} {\Psi}_{2,4} - {\bar{\Psi}}_{2,2} - {\bar{\Psi}}_{2,4} {\bar{\Psi}}_{2,1} {\Psi}_{2,3} ] |{\Omega}_1 \rangle, \nonumber \\
\frac{\partial}{\partial M^2} P_+ ( ^{\star} {\bar{\Psi}}_{2,3} |0 \rangle) &= \frac{1}{2M} [ -{\bar{\Psi}}_{2,3} {\bar{\Psi}}_{2,1} {\Psi}_{2,2} + {\bar{\Psi}}_{2,2} {\bar{\Psi}}_{2,3} {\Psi}_{2,1} - {\bar{\Psi}}_{2,4} ] |{\Omega}_1 \rangle,
\ee
\be
\frac{\partial}{\partial M^3} P_- {\bar{\Psi}}_{2,2} |0 \rangle &= \frac{i}{2M} [ -{\bar{\Psi}}_{2,1} {\Psi}_{2,4} {\Psi}_{2,2} + {\bar{\Psi}}_{2,2} {\Psi}_{2,4} {\Psi}_{2,1} + {\Psi}_{2,3} ] |{\Omega}_1 \rangle, \nonumber \\
\frac{\partial}{\partial M^3} P_- {\bar{\Psi}}_{2,4} |0 \rangle &= \frac{i}{2M} [ {\bar{\Psi}}_{2,3} {\Psi}_{2,4} {\Psi}_{2,2} + {\Psi}_{2,1} - {\bar{\Psi}}_{2,4} {\Psi}_{2,3} {\Psi}_{2,2}]|{\Omega}_1 \rangle, \nonumber \\
\frac{\partial}{\partial M^3} P_+ ( ^{\star} {\bar{\Psi}}_{2,1} |0 \rangle) &= \frac{i}{2M} [ -{\bar{\Psi}}_{2,1} {\bar{\Psi}}_{2,3} {\Psi}_{2,4} - {\bar{\Psi}}_{2,2} + {\bar{\Psi}}_{2,4} {\bar{\Psi}}_{2,1} {\Psi}_{2,3} ] |{\Omega}_1 \rangle, \nonumber \\
\frac{\partial}{\partial M^3} P_+ ( ^{\star} {\bar{\Psi}}_{2,3} |0 \rangle) &= \frac{i}{2M} [ {\bar{\Psi}}_{2,3} {\bar{\Psi}}_{2,1} {\Psi}_{2,2} - {\bar{\Psi}}_{2,2} {\bar{\Psi}}_{2,3} {\Psi}_{2,1} + {\bar{\Psi}}_{2,4} ] |{\Omega}_1 \rangle,
\ee
\be
\frac{\partial}{\partial M^4} P_- {\bar{\Psi}}_{2,2} |0 \rangle &= \frac{i}{2M} [ -{\bar{\Psi}}_{2,3} {\Psi}_{2,4} {\Psi}_{2,2} - {\Psi}_{2,1} + {\bar{\Psi}}_{2,2} {\Psi}_{2,4} {\Psi}_{2,3} ] |{\Omega}_1 \rangle, \nonumber \\
\frac{\partial}{\partial M^4} P_- {\bar{\Psi}}_{2,4} |0 \rangle &= \frac{i}{2M} [ -{\bar{\Psi}}_{2,1} {\Psi}_{2,4} {\Psi}_{2,2} - {\bar{\Psi}}_{2,4} {\Psi}_{2,2} {\Psi}_{2,1} + {\Psi}_{2,3} ]|{\Omega}_1 \rangle, \nonumber \\
\frac{\partial}{\partial M^4} P_+ ( ^{\star} {\bar{\Psi}}_{2,1} |0 \rangle) &= \frac{i}{2M} [ -{\bar{\Psi}}_{2,3} {\bar{\Psi}}_{2,1} {\Psi}_{2,2} - {\bar{\Psi}}_{2,4} + {\bar{\Psi}}_{2,1} {\bar{\Psi}}_{2,2} {\Psi}_{2,3} ] |{\Omega}_1 \rangle, \nonumber \\
\frac{\partial}{\partial M^4} P_+ ( ^{\star} {\bar{\Psi}}_{2,3} |0 \rangle) &= \frac{i}{2M} [ -{\bar{\Psi}}_{2,1} {\bar{\Psi}}_{2,3} {\Psi}_{2,4} + {\bar{\Psi}}_{2,3} {\bar{\Psi}}_{2,4} {\Psi}_{2,1} - {\bar{\Psi}}_{2,2} ] |{\Omega}_1 \rangle,
\ee
and finally the $M_5$ derivatives
\be
\frac{\partial}{\partial M^5} P_- {\bar{\Psi}}_{2,2} |0 \rangle &= -\frac{2}{M} {\Psi}_{2,4} |{\Omega}_1 \rangle, \nonumber \\
\frac{\partial}{\partial M^5} P_- {\bar{\Psi}}_{2,4} |0 \rangle &= -\frac{2}{M} {\Psi}_{2,2}|{\Omega}_1 \rangle , \nonumber \\
\frac{\partial}{\partial M^5} P_+ ( ^{\star} {\bar{\Psi}}_{2,1} |0 \rangle) &= -\frac{2}{M} {\bar{\Psi}}_{2,1} |{\Omega}_1 \rangle , \nonumber \\
\frac{\partial}{\partial M^5} P_+ ( ^{\star} {\bar{\Psi}}_{2,3} |0 \rangle) &= -\frac{2}{M} {\bar{\Psi}}_{2,3} |{\Omega}_1 \rangle .
\ee
\para
We wish to compute the connection for the states living in the ${\Psi}_{i \alpha}|0\rangle$-sector where $i=1,2$ and $\alpha=1,\dots,4$ which is given schematically by
\be
{\omega}_{\mu} = \left(\begin{array}{cc}
\langle 2 | {\partial}_{\mu} | 2 \rangle & \langle 1 | {\partial}_{\mu} | 2 \rangle \\
\langle 2 | {\partial}_{\mu} | 1 \rangle & \langle 1 | {\partial}_{\mu} | 1 \rangle
\end{array}\right),
\ee
where the state $| 1F_1 \rangle$ is the pair of 1-fermion first-excited states above Vacuum $1$.
\para
In the 3-fermi sector $^{\star} {\Psi}_{1 \alpha}$, we calculate similar overlaps, and arrange the connection as
\be
{^{\star} \Sigma}_{\mu} = \left(\begin{array}{cc}
\langle ^{\star} 1 | {\partial}_{\mu} | ^{\star} 1 \rangle & \langle ^{\star} 1 | {\partial}_{\mu} | ^{\star} 2 \rangle \\
\langle ^{\star} 2 | {\partial}_{\mu} | ^{\star} 1 \rangle & \langle ^{\star} 2 | {\partial}_{\mu} | ^{\star} 2 \rangle
\end{array}\right).
\ee
Calculating these overlaps explicitly gives the Berry phase connection components in the main body of the text.

\newpage
\section*{Appendix B: Instantons in $(4,4)$ Quantum Mechanics}
\setcounter{section}{2}

Having calculated the form of correlators that can get instanton corrections in the $(4,4)$ quantum mechanics, we now need to consider the form of the instanton measure for these contributions, in order that we may find explicitly the function $f(Mr)$. We decompose the measure into three parts
\be
\int \, d{\mu}_{inst} = e^{-S_{inst}} \int \, d{\mu}_B \int \, d{\mu}_F \; \text{dets},
\ee
and treat each factor in turn.

\vspace{2ex}

\subsection*{Bosonic Measure $d{\mu}_B$}

The bosonic measure arises from the zero-mode solutions to the Bogomolnyi equations (\ref{bogeqn}). Since we have the same Bogomolnyi equations as in the $(2,2)$ case of \cite{geophase}, we expect that the bosonic measure for the problem is
\be
\int \, d{\mu}_B = \int \, \frac{dT}{\sqrt{2 \pi}} \sqrt{g_{TT}} \int \, \frac{d \theta}{\sqrt{2 \pi}} \sqrt{g_{\theta \theta}} = r \int \, dT .
\ee

\vspace{2ex}

\subsection*{Fermionic Measure $d{\mu}_F$}

The fermions in our problem are now more complicated, since each bosonic zero-mode is now related to two fermionic zero-modes by the unbroken supersymmetry generators in the kink background. We therefore expect some alteration to the total fermionic measure.

As previously implied, we still have the $(2,2)$ zero-modes, and so we still have the broken susy transformations
\be
{\lambda}_- &= -i ({\partial}_{\tau} X) {\epsilon}_- \;\;\; , \;\;\; {\bar{\psi}}_{i+} = \sqrt{2} {\cal{D}}_{\tau} {\phi}_i^{\dagger} {\epsilon}_- , \nonumber \\
{\bar{\lambda}}_+ &= i({\partial}_{\tau} X) {\bar{\epsilon}}_+ \;\;\; , \;\;\; {\psi}_{i-} = -\sqrt{2} {\cal{D}}_{\tau}
{\phi}_i {\bar{\epsilon}}_+ ,
\ee
but we also have broken susy transformations for the other set of zero modes ${\tilde{\epsilon}}_+$ and ${\bar{\tilde{\epsilon}}}_-$. However, we need not calculate these explicitly, since the second set of zero modes have the same equations of motion, and can simply state that the fermionic measure is the square of what we found in the $(2,2)$ case;
\be
\int \, d{\mu}_F = \int  \, d{\epsilon}_- d{\bar{\epsilon}}_+ d{\tilde{\epsilon}}_+ d{\bar{\tilde{\epsilon}}}_- \; [J_{(2,2)}]^{-2} = \frac{1}{4M^2 r^2} \int \, \, d{\epsilon}_- d{\bar{\epsilon}}_+ d{\tilde{\epsilon}}_+ d{\bar{\tilde{\epsilon}}}_- .
\ee

\vspace{2ex}

\subsection*{1-Loop Determinants}

Finally, we need to be able to do the $1$-loop Gaussian integrals around the background of the kink in order to know the contribution of the non-zero modes to the instanton measure. For the $(2,2)$ quantum mechanics, there is incomplete cancelation between the contributions from the bosons and the fermions on account of a spectral asymmetry of the Dirac operator. For the more highly supersymmetric $(4,4)$ theory, we expect that there is total cancelation between the modes, leading to a determinant ratio that is precisely unity. We can do better, however, and give a physicist ``proof'' that this is the case.

We work with the fermions first, and treat them as quantum fluctuations, since the kink solution has no fermionic excitations. Performing the Gaussian integrals, we get a simple contribution that is just the square of what we found for the $(2,2)$ case
\be
{\Gamma}_F = \biggl[ \frac{ \det ({\Delta}^{\dagger}) {\det}^{\prime} ( \Delta)}{\det({\Delta}_0^{\dagger} {\Delta}_0)} {\biggr]}^2 = \biggl[ \frac{ \det( \Delta {\Delta}^{\dagger} ) {\det}^{\prime} ({\Delta}^{\dagger} \Delta)}{{\det}^2({\Delta}_0^{\dagger} {\Delta}_0)} \biggr] .
\ee
\para
Next we consider the Bosonic action (\ref{bosac}), and expand all fields about their classical expectation values as follows
\be
{\phi}_i &=& {\phi}_i^{cl} + \delta{\phi}_i , \nonumber \\
{\tilde{\phi}}_i &=& \delta {\tilde{\phi}}_i , \nonumber \\
{\vec{X}} &=& (\delta X_1,\delta X_2,\delta X_3,\delta X_4,X_5^{cl} + \delta X_5) , \nonumber \\
A_0 &=& \delta A_0 .
\ee

Because their classical expectation value is zero, the fields ${\tilde{\phi}}_i$ completely decouple from the other
fields at one-loop, and so their small fluctuation operator is particularly simple
\be
{\Delta}_{\tilde{\phi}_i} = -{\cal{D}}_{\tau}^2 + |X-M_i|^2 ,
\ee
and so they contribute
\be
{\Gamma}_{\tilde{\phi}} = \left( \frac{\det({\Delta}_{\tilde{\phi}_i})}{\det({\Delta}_{\tilde{\phi}_i, 0})} \right).
\ee

We get two copies of the determinant for the fields $(\delta X_1,\delta X_2)$, so the $X_i \; i=1,\dots,4$ excitations give determinants
\be
{\Gamma}_{X_i} = \left( \frac{ \det(-{\partial}_{\tau}^2 + 2g^2 r)}{\det (-{\partial}_{\tau}^2 + 2g^2 |{\phi}_i|^2)} \right)^2 ,
\ee
one of which cancels with the ghosts that arise from fixing the gauge to $A_0 = 0$; they give a one-loop contribution
\be
{\Gamma}_{ghost} =  \frac{ \det(-{\partial}_{\tau}^2 + 2g^2 |{\phi}_i|^2)}{\det(-{\partial}_{\tau}^2 + 2g^2 r)} .
\ee
Finally, we get a small fluctuation operator for the fields $(\delta X_5, \delta {\phi}_i)$ that corresponds with the fermionic operator ${\Delta}^{\dagger} \Delta$, so that the total contribution from the Bosons is
\be
{\Gamma}_B &=&  {\Gamma}_{\tilde{\phi}} {\Gamma}_{X_i} {\Gamma}_{ghost} {\Gamma}_{(X_5,{\phi}_i)} \nonumber \\
&=& \left[ \frac{ \det ( -{\partial}_{\tau}^2 + 2g^2 r) \det( {\Delta}_0^{\dagger} {\Delta}_0) \det({\Delta}_{\tilde{\phi}_i, 0})}{ \det ( -{\partial}_{\tau}^2 + 2g^2 |{\phi}_i|^2) {\det}^{\prime}({\Delta}^{\dagger} \Delta ) \det(-D_{\tau}^2 + |X_5 - M_i|^2)} \right] .
\ee

At first sight this looks rather a mess, but when we put it together with the fermionic contribution, a series of magic cancelations occur; we are left with
\be
\Gamma = \left[ \frac{ \det({\Delta}{\Delta}^{\dagger})}{ \det(-{\partial}_{\tau}^2 + 2g^2|{\phi}_i|^2 ) \det ( -{\cal{D}}_{\tau}^2 + |X-M_i|^2)} \right] ,
\ee
but after a little algebra, we find that ${\Delta}{\Delta}^{\dagger} = \text{diag}(-{\partial}_{\tau}^2 +2g^2 |{\phi}_i|^2, -{\cal{D}}_{\tau}^2 + |X-M_i|^2)$, and so we see that
\be
\Gamma = 1.
\ee
The $1$-loop determinants are indeed trivial as anticipated. We can now put everything together to find our full instanton measure. It is
\be
\int \, d{\mu}_{inst} = \frac{e^{-2Mr}}{4M^2 r} \int \, dT \int \, d{\epsilon}_- d{\bar{\epsilon}}_+ d{\tilde{\epsilon}}_+ d{\bar{\tilde{\epsilon}}}_- .
\ee

Finally, we need the asymptotic behaviour of the fermions in $f$, which comes from solving the Bogomolnyi equations (\ref{bogeqn}) in $A_0=0$ gauge to give ~\cite{domwall}
\be
{\phi}_1 = \frac{\sqrt{r} e^{M \tau}}{\sqrt{e^{2M \tau} + e^{-2M \tau}}} \;\;\; , \;\;\;
{\phi}_2 = \frac{\sqrt{r} e^{-M \tau}}{\sqrt{e^{2M \tau} + e^{-2M \tau}}}.
\ee
The applying the fermionic forms of the Bogomolnyi equations, we find that
\be
{\psi}_1 \sim {\tilde{\psi}}_1 \sim \sqrt{2} {\mathcal{D}}_{\tau} {\phi}_1 , \nonumber \\
{\psi}_2 \sim {\tilde{\psi}}_2 \sim \sqrt{2} {\mathcal{D}}_{\tau} {\phi}_2 ,
\ee
which tells us the explicit form of the function $K$
\be
K = (\sqrt{2} {\mathcal{D}}_{\tau} {\phi}_2)^3 (\sqrt{2} {\mathcal{D}}_{\tau} {\phi}_1),
\ee
and so putting everything together and doing the integral, we find that
\be
f(Mr) = \frac{3 \pi Mr}{16} e^{-2Mr}.
\ee

However, note that this is merely the leading order part of the instanton correction, presumably higher order terms are generated by higher-loop effects in the instanton background, and there is no reason to believe that these vanish.

\subsection*{Acknowledgements}

It is a pleasure to acknowledge the help of David Tong at all stages of this work. I would also like to thank Julian Sonner for useful comments and discussions. I have been supported by Trinity College, Cambridge and EPSRC.

 \end{document}